\documentclass[pre,twocolumn]{revtex4}
\usepackage{epsfig,amssymb,amsmath,latexsym,color,pifont,graphicx}
\newcommand{\e}{\varepsilon}
\newcommand{\p}{\varphi}
\newcommand{\s}{\sigma}
\newcommand{\om}{\omega}
\newcommand{\er}{\eqref}
\newcommand{\RR}{R_0}
\newcommand{\PP}{\Phi_0}
\newcommand{\ZZ}{Z_0}
\newcommand{\bbR}{\bar{R}}
\newcommand{\bbP}{\bar{\Phi}}
\newcommand{\bbZ}{\bar{Z}}
\newcommand{\bbRR}{\bar{R}_0}
\newcommand{\bbPP}{\bar{\Phi}_0}
\newcommand{\bbZZ}{\bar{Z}_0}

\renewcommand\Im{\operatorname{Im}}

\begin{document}

\title{Phase resetting of collective rhythm in ensembles of oscillators}

\begin{abstract}
Phase resetting curves characterize the way a system with a collective periodic behavior responds to perturbations. We consider globally coupled ensembles of Sakaguchi-Kuramoto oscillators, and use the Ott-Antonsen theory of ensemble evolution to derive the analytical phase resetting equations. We show the final phase reset value to be composed of two parts: an immediate phase reset directly caused by the perturbation, and the dynamical phase reset resulting from the relaxation of the perturbed system back to its dynamical equilibrium. Analytical, semi-analytical and numerical approximations of the final phase resetting curve are constructed. We support our findings with extensive numerical evidence involving identical and non-identical oscillators. The validity of our theory is discussed in the context of large ensembles approximating the thermodynamic limit.
\end{abstract}

\author{Zoran Levnaji\'c}
\author{Arkady Pikovsky}
\affiliation{Department of Physics and Astronomy, University of Potsdam, 14476 Potsdam, Germany}
\date{\today}

\maketitle

%------------------------------------------------------------------------------------------------------------------------------------------------------------------------------------------

\section{Introduction}  \label{Sec-Introduction}

The behavior of many biological systems displays rhythmic patterns that can be globally described through a periodic \textit{phase} variable~\cite{winfree-book}. Examples are found on a variety of time-scales and levels of complexity: pulsating neurons, circadian clocks in living beings, seasonal dynamics etc.~\cite{winfree-book,tass-book,granada}. The response of a periodic system to stimuli can be characterized by measuring the phase shift occurring as a result of a stimulus, i.e. by quantifying the difference between the collective phases of the perturbed and the original system. The phase shift's dependence on the phase value at which the perturbation occurred is termed \textit{phase resetting curve} (or phase response curve, PRC), and is an inherent characteristic of any oscillatory system~\cite{kuramoto-book,ermentrout-96}. Study and applications of PRCs are currently receiving a growing attention through their theoretical and experimental aspects~\cite{tass-book,granada}.

Periodic biological systems are often composed of many interacting units, whose collective behavior emerges from the dynamical properties of the single units~\cite{tass-book}. The collective (macroscopic) phase response of an empirical oscillator is generated from stimuli that act on the level of individual (microscopic) units, and composed of their individual phase responses.

In the context of neural oscillations various methods have been developed for the empirical measurement of PRCs, both \textit{in vivo} and \textit{in vitro}. The methods depend on the type/size of the neural ensemble, and are typically involving electric stimulation of neurons that induce a measurable phase shift in their spiking behavior~\cite{tass-book,galan-ermen,tateno}. Recently, the epileptiform activity in rats was characterized through experimental PRCs, obtained by stimulating thalamus and cortex of epileptic animals during spike and wave seizures~\cite{ramon}. Investigation of PRCs is relevant for understanding the interaction properties of the neural networks, such as their stability~\cite{tateno}, or synchronization and clustering~\cite{achuthan1}. The assumption of weak inter-neuron coupling was experimentally tested via infinitesimal PRCs~\cite{preyer}. A new technique of PRC construction from data, based on weighted spike-triggered averaging was recently proposed~\cite{ota}. While mostly studied in the domain of neurons, PRCs are also explored in other natural rhythmic systems, that can be modeled as ensembles of many oscillatory elements. Empirical and theoretical examination was thus conducted in coupled circadian clocks of insects~\cite{petri}, and the periodically driven saline oscillator~\cite{gonzo}.

Due to its global periodic behavior, dynamics of neurons/oscillators and the resulting PRCs can be easily studied analytically or modeled computationally~\cite{tass-book,galan-ermen,jeff}. On the level of few neurons, extensive analytical results including bifurcation diagrams are available for various neuron models and types of interaction~\cite{tateno,ramon,jeff,croisier}. On the other hand, global cooperative behavior in large neuron ensembles is typically examined numerically by modeling single elements as simple phase-oscillators. In this context, Kuramoto-type oscillators~\cite{kuramoto-book} are usually employed for their elegant and useful synchronization properties~\cite{prk-book}, that are well understood for various topologically and structurally different networks~\cite{acebron,arenas}. A recent study of the phase response of stochastic oscillators revealed the interplay between microscopic and macroscopic collective phase sensitivity, for ensembles of globally coupled~\cite{kawamura}, and network coupled elements~\cite{kori}. Phase resetting in globally coupled ensembles of oscillators was investigated in relation to the symmetry properties of the coupling function~\cite{ko-ermen}, emphasizing the peculiarity of non-odd coupling functional forms. PRCs were also constructed for models involving pulse-coupled neurons, providing insights into their interaction details and collective dynamics at small~\cite{talathi} and large scales~\cite{achuthan2}. Phase response behavior of complex oscillator networks, besides giving insights into their interaction patterns~\cite{achuthan1,kralemann}, can also assist in revealing the details of their network connectivity~\cite{timme}.

While the PRCs of a single (or few) oscillators are very well understood~\cite{kuramoto-book,galan-ermen,tateno,croisier}, the results for ensembles of oscillators are still rather scarce and often limited to numerical investigations only~\cite{granada,kori,ko-ermen}. Namely, the difficulty here is that there are two contributions to the phase resetting. Apart from the phase shift immediately caused by the perturbation, the relaxation of the ensemble into a stationary state may induce an additional (potentially significant) phase shift, since any finite stimulus in general perturbs the system out of its equilibrium. 

In this paper we consider a globally coupled ensemble of Sakaguchi-Kuramoto oscillators~\cite{sakaguchi}, interacting via a non-odd coupling function. The system is perturbed from its dynamical equilibrium by applying the kick (of controllable form and strength) to the individual oscillators. We employ the Watanabe-Strogatz ansatz~\cite{watanabe} in the recently proposed Ott-Antonsen (OA) formulation~\cite{OA}, and calculate analytically the effect of the relaxation of system's collective phase. As we show, an additional phase reset is induced, and its properties can be obtained from the ensemble's state measured immediately after the perturbation. We confirm our theoretical results through a series of numerical simulations, considering both identical and non-identical oscillators. The extent of our theory is illustrated by considering the phase-resetting in an ensemble of Stuart-Landau oscillators.

In opposition to~\cite{kawamura,kori}, here we consider deterministic oscillators and perturbations of arbitrary strength (not infinitesimal), thus allowing the exploration of the phase reset's time evolution.

This paper is organized as follows: in Section~\ref{Sec-Formulation} we formulate our model of kicked oscillator ensemble, define the concept of phase resetting and describe our implementation of perturbation. In Section~\ref{Sec-AnalyticalZ0} we show some analytical results regarding calculation of the immediate PRC. In Section~\ref{Sec-Time-evolution} we derive our main analytical result based on OA theory, for a general case of non-identical oscillators. We test our theory in Section~\ref{Sec-Numerics} by exposing our numerical results for various oscillator ensembles and types. We discuss our results and conclude in Section~\ref{Sec-Conclusions}.

%------------------------------------------------------------------------------------------------------------------------------------------------------------------------------------------

\section{Formulation of our model}  \label{Sec-Formulation}

In this Section we construct our system of globally coupled oscillators and define the concept of phase resetting curve for the ensemble. We describe the implementation of the perturbation (kick) in response to which 
the PRC is measured.

\subsection{Ensemble and its description}
We consider an ensemble of $N$ oscillators characterized by their natural frequencies $\om_k$, whose dynamical states are defined by their phase values $\p_k \in [0,2\pi)$. Oscillators are globally coupled using Sakaguchi-Kuramoto scheme with the interaction strength $\e>0$ and phase shift $\beta \in (-\frac{\pi}{2},\frac{\pi}{2})$: 
\begin{equation} \dot{\p_k} = \om_k + \frac{\e}{N} \sum_{j=1,N} \sin (\p_j - \p_k + \beta) + q(\p_k;\vec{a}_k) \; \delta(t-T)\;. \label{eq-KSmodel} \end{equation}
The oscillator $\p_k$ undergoes a kick at time $t=T$, whose properties are given through the function $q$, which in general depends on the phase $\p_k$ and a set of parameters summarized in $\vec{a}_k$. We take $q$ to have a simple form as follows:
\begin{equation} q(\p;\vec{a}) =  q(\p;A,\alpha) = A \sin (\p + \alpha) \label{eq-kick} \end{equation}
where $A$ is the kicking strength, and $\alpha$ a phase shift parameter. The collective dynamical state of the ensemble is quantified through the \textit{complex mean field} (complex order parameter) $Z$ defined as:
\[ Z = \frac{1}{N} \sum_{k=1,N} e^{i \p_k} = \langle e^{i \p} \rangle  = R e^{i \Phi} \]
where $\langle \cdot \rangle$ stays for the average over ensemble. We call its absolute value $R=|Z|$ the collective radius, and its argument $\Phi = \arg Z$ the collective phase. The ensemble's PRC is defined with respect to the collective phase $\Phi$.

We now construct an alternative formulation of our system's equations by transforming the phase $\p$ into a unitary complex variable $\s$:
\[ \s = e^{i\p} \;. \]
Thus, the time evolution of our kicked system given by Eq.~\er{eq-KSmodel} now reads:
\begin{equation} \dot \s_k = i \s_k \;  \Im \Big( i  \om_k + \e e^{i\beta} Z  \s_k^* + r (\s_k;\vec{a}_k) \; \delta(t-T) \Big) \label{eq-KSmodel-s} \end{equation}
where $\s^*$ is the complex conjugate of $\s$. The functional expression for the kick $r$ becomes:
\begin{equation} r(\s;\vec{a}) = r(\s;A,\lambda) = A \lambda \s   \label{eq-kick-s} \end{equation}
where $\lambda = e^{i\alpha}$. The complex mean field $Z$ reads:
\[ Z = \frac{1}{N} \sum_{k=1,N} \s_k = \langle \s \rangle \; .  \]
We term this approach \textit{unitary complex representation}, in contrast to previously exposed \textit{phase representation}. It transforms the trigonometric expressions in our equations into the algebraic ones, which will be of importance in our analytical studies that follow. Throughout this paper we will be interchangeably using both representations.

Two types of phase-oscillator ensembles in respect to the frequency distributions are considered:\\[0.1cm]
(\textit{i}) \textit{identical oscillators} \,\, $\om_k=\om$, whose final dynamical state is the full synchronization $\p_k=\Phi$ (regardless of $N$, $\e>0$ and $\beta \in (-\frac{\pi}{2},\frac{\pi}{2})$). 
We will examine examples of small ensembles $N < \infty$, and large ensembles approximating thermodynamic limit $N \rightarrow \infty$.\\[0.1cm]
(\textit{ii}) \textit{non-identical oscillators} \,\, with Lorentzian frequency distribution $g$, characterized by a mean $\tilde{\omega}$ and a width $\gamma \ge 0$:
\begin{equation}   g(\omega) = \dfrac{1}{\pi}  \dfrac{\gamma}{(\omega - \tilde{\omega})^2 + \gamma^2} \;. \label{eq-lorentzian}  \end{equation}
After relaxation this ensemble is partially synchronized (depending on $\beta$ and $\gamma$). Here we consider only the thermodynamic limit $N \rightarrow \infty$. The case of identical oscillators is obtained for 
$\gamma \rightarrow 0^+$.

\subsection{General definition of PRC for the ensemble}
Consider an ensemble of $N>1$ oscillators governed by Eq.~\er{eq-KSmodel} or \er{eq-KSmodel-s} whose collective dynamical state is given by the complex mean field $Z$. We compare two realizations of the ensemble, termed ``original'' and ``kicked''  system, with time evolutions described by $Z(t)$ and $\bbZ(t)$ respectively (all quantities related to the kicked system are denoted with bar). We assume the kicked system to be created at time $t=T=0$ in its initial state $\bbZ(0)=\langle \bar\s(0)\rangle=\bbZZ=\langle \bar\s_0\rangle $ as the consequence of the kick acting on the original system, which at time $t=0$ is in its stationary state $Z(0)=Z_0$:
\[ Z_0  \stackrel{\mbox{kick}}{\xrightarrow{\hspace*{0.8cm}}} \bbZZ \qquad \mbox{or} \qquad (\PP, R_0)\stackrel{\mbox{kick}}{\xrightarrow{\hspace*{0.8cm}}}  (\bbPP, \bbRR)\;.  \] 
After the kick, two systems independently continue their time evolutions -- original system starting from $Z_0$ and kicked system starting from $\bbZ_0$.

The phase resetting curve is defined as the difference between the collective phases of two systems:
\[ \Delta = \bbP - \Phi =  \arg \frac{\bbZ}{Z} \]
as function of the collective phase value (of the original system) at which the kick occurred $\Delta = \Delta (\PP)$. We differ between two possible PRCs, depending on the time of phase reset measurement:
\begin{itemize}
\item We call \textit{immediate phase resetting curve} (pPRC) $\Delta_0 = \Delta_0 (\PP)$ the phase shift value observed immediately after the resetting, i.e. as the immediate consequence of the kicking:
\begin{equation} \Delta_0 = \bbPP - \PP  = \arg \frac{\bbZZ}{Z_0}  = \arg  \frac{\langle \bar\s_0 \rangle}{\langle \s_0 \rangle}    \label{eq-defIPRC} \end{equation}
      (we chose the abbreviation pPRC for ``prompt PRC'', since the abbreviation IPRC is in use for ``infinitesimal PRC''~\cite{preyer}).
\item In contrast to pPRC, we term \textit{final phase resetting curve} (fPRC) $\Delta_\infty = \Delta_\infty (\PP)$ the eventual phase reset value, measured when the kicked system relaxes to its final stationary state: 
\[  \Delta_\infty = \lim_{t \rightarrow \infty} \big(\bbP - \Phi\big)(t)  =  \lim_{t \rightarrow \infty}  \arg \frac{\bbZ(t)}{Z(t)}\;. \]
\end{itemize}

What we show in this paper, is that the curves $\Delta_0$ and $\Delta_\infty$ are in general different. We illustrate this on a simple example. Consider $N=10$ identical oscillators whose collective dynamics is given by Eq.~\er{eq-KSmodel}. They define the original system, synchronized ($R_0=1$) at the phase value $\PP$ where the kick occurs. The kick acts asymmetrically: 5 out of 10 oscillators are kicked with a uniform strength $A$. The kicked system is created in its initial state $\bbZZ = \bbRR e^{i\bbPP}$ with $\bbRR<1$, while the original system continues to rotate with radius $R(t)=1$. 
\begin{figure}[!ht]  \begin{center}
       \includegraphics[height=2.6in,width=3.3in]{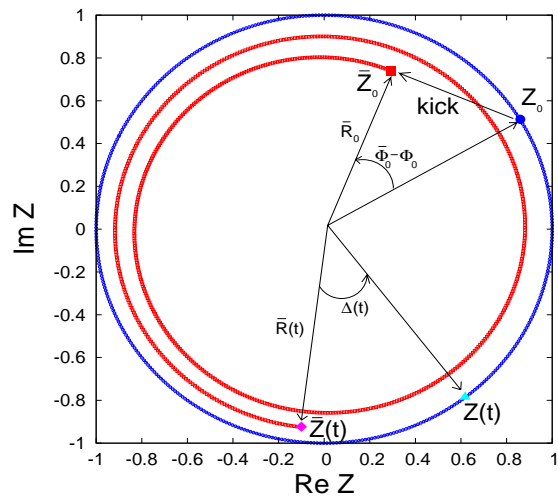}
  \caption{(Color online) Ensemble of $N=10$ identical oscillators $\om_k=\om=1$ undergoing a kick at collective state $\ZZ = e^{i\PP}$ (large dot). 5 oscillators are kicked with strength $A=-0.25$ (other oscillators are left unperturbed), for $\e=0.15$ and $\beta=\frac{\pi}{3}$. Kicked system starts from its initial state $\bbZZ = \bbRR e^{i\bbPP}$ (large square). Time-evolutions of both complex mean fields are shown for comparison: original system $Z(t)$ (dark/blue line) and kicked system $\bbZ (t)$ (light/red line). The values of $\bbPP - \PP = \Delta_0$, and $\Delta (t)$ are shown.}  \label{fig-1} 
\end{center} \end{figure}
In Fig.~\ref{fig-1} we show the evolution of the original system's complex mean field (blue), in comparison with the one for the kicked system (red). The value of pPRC $\Delta_0 = \bbPP - \PP$ is clearly visible and can be calculated from Eq.~\er{eq-defIPRC} (we discuss this in the next Section). $\Delta_\infty$ is measured once the kicked system reaches the synchronized state ($\bbR(t) \rightarrow 1$), i.e. it is the limit value of $\Delta (t)$. As we show in what follows, the system's transient desynchronization induces different rotation speeds of the two collective phases which ultimately leads to a discrepancy between $\Delta_0$ and  $\Delta_\infty$.

\subsection{Implementation of the kick}
We now construct the transformation that maps the phase of an oscillator from its pre-kick to its post-kick value:
\[  \p_0  \stackrel{\mbox{kick}}{\xrightarrow{\hspace*{0.8cm}}}  {\bar \p}_0  \]
in accordance with Eq.~\er{eq-kick} and \er{eq-kick-s}. For $N=1$ and $T=0$ the Eq.~\er{eq-KSmodel} reduces to: 
\[ \dot{\p} = \om +  A \sin (\p + \alpha)\; \delta(t)  \;. \]
We solve this equation by assuming that at the time $t=0$ the frequency term is negligible. This gives:
\[  Q ({\bar \p}_0) - Q(\p_0) = 1  \qquad \mbox{where} \qquad Q = \int \frac{d\p}{A \sin (\p + \alpha)}\;, \]
which yields the relation
\[  \tan \frac{{\bar \p}_0 + \alpha}{2} = e^A \tan \frac{\p_0 + \alpha}{2}\;,  \] 
from which the final transformation is obtained:
\begin{equation} {\bar \p}_0 \; = \; 2  \arctan \left( e^{A} \; \tan \frac{\p_0 + \alpha}{2}\; \right)  - \alpha\;.  \label{eq-kickjump}  \end{equation}
We use the Eq.~\er{eq-kickjump} for shifting the phase value of each oscillator in the ensemble as the consequence of the kick. Also, this expression automatically gives the PRC of an isolated oscillator $\Delta^S$ in response to the kick Eq.~\er{eq-kick}:
\[ \Delta^S =  {\bar \p} (0^+) - \p (0^-) = {\bar \p}_0 - \p_0 \;.\] 
Since the properties of the time evolution of an isolated oscillator are not altered by the perturbation (no relaxation back to stationary state), its pPRC is always identical to its fPRC.

Alternatively, the kick transformation Eq.~\er{eq-kickjump} can be formulated in terms of the unitary complex representation. Proceeding equivalently as above, we obtain:
\[ {\bar \s}_0  \; = \; \frac{\lambda\s_0 - b}{\lambda - \lambda^2 b \s_0} \;, \]
where $b = \frac{e^A - 1}{e^A + 1} = \tanh \frac{A}{2}$. Furthermore, if we introduce the complex parameter $\eta = b \lambda$, the transformation above can be further simplified as:
\begin{equation} {\bar \s}_0  \; = \; \frac{\s_0 - \eta^*}{1 - \eta \s_0}\;.   \label{eq-kickjump-s}  \end{equation}
The PRC for a single oscillator $\Delta^S$ now reads:
\[ \Delta^S =  \arg \frac{{\bar \s}_0}{\s_0} = \arg \frac{1 - (\eta{\s_0})^*}{1 - \eta \s_0} \;. \] 
The kick transformation given by Eq.~\er{eq-kickjump} or \er{eq-kickjump-s} can be used together with Eq.~\er{eq-defIPRC} to analytically calculate $\bbZZ$ (and hence pPRC) for deterministic kicks.

%------------------------------------------------------------------------------------------------------------------------------------------------------------------------------------------

\section{Analytical Calculation of $\bbZZ$} \label{Sec-AnalyticalZ0}

In this Section we discuss the calculation of $\bbZZ$ in the thermodynamic limit for the cases when the kicking parameters are picked from a given probability distribution. For a general ensemble with frequency distribution $g$, the complex mean field reads:
\[ Z  =  \int d\om d\p \;  g(\om) U(\p,\om) e^{i\p}  =  \int d\om d\s \; g(\om) W(\s,\om) \s \] 
where $U(\p,\om)$ is the fraction of oscillators having frequency $\om$ and phase $\p$ (respectively, $W(\s,\om)$ for $\om$ and $\s$). After the kick, the complex mean field $\bbZZ = \langle {\bar \s}_0 \rangle$ is:
\begin{equation} \begin{array}{ll}
 \bbZZ & =  \int dA\, d\alpha\, d\om\, d\p_0 \; g(\om) \rho (\om;A,\alpha) U (\p_0,\om) e^{i {\bar \p}_0(\p_0)}  \\[0.2cm]
       & =  \int dA\, d\alpha\, d\om\, d\s_0 \; g(\om) \rho (\om;A,\alpha) W (\s_0,\om) {\bar \s}_0 (\s_0) 
\end{array}  \label{eq-bbZZ-int}  \end{equation}
where $\rho$ stays for a given distribution of kicking parameters $A$ and $\alpha$, that may depend on $\om$. Thus, $\bbZZ$ can in principle always be calculated if the distributions $U$ (or $W$) and $\rho$ are known, and the pPRC can be obtained through Eq.~\er{eq-defIPRC} (although, the integral Eq.~\er{eq-bbZZ-int} can often be solved only numerically). As our first numerical example, we study in Section~\ref{Sec-Numerics} an ensemble of identical oscillators with kicking strengths $A$ picked from a uniform distribution $A \in [-A_0,A_0]$. We compare the simulation results with the 'analytical' curves obtained by calculating integrals Eq.~\er{eq-bbZZ-int} numerically.

Alternatively, one can expand the expression Eq.~\er{eq-kickjump-s} in power series to obtain:
\[ \begin{array}{lll}
   \bbZZ & = \Big\langle (\s_0 - \eta^*) \sum_{n=0}^{\infty}  (\eta \s_0)^n \Big\rangle & \\[0.2cm]
   &  = \Big\langle \s_0 \sum_{n=0}^{\infty}  (\eta \s_0)^n \Big\rangle   -  \Big\langle \eta^* \sum_{n=0}^{\infty}  (\eta \s_0)^n  \Big\rangle \;, &
\end{array} \]
valid for $|\eta \s_0|<1$ that is always verified in our case. We employ the expression above to directly calculate $\bbZZ$ for other two examples numerically studied in Section~\ref{Sec-Numerics}.

As the first of them, we consider identical oscillators with a unique kicking strength $b=\tanh\frac{A}{2}$, and phase shifts $\alpha$ picked from the following distribution:
\begin{equation} \rho (\alpha) = \frac{1}{2\pi} \left(1 + 2 S \cos(\alpha - \tilde{\alpha}) \right) \;, \label{eq-walpha} \end{equation}
where $S < \frac{1}{2}$ and $\tilde{\alpha} \in (-\frac{\pi}{2},\frac{\pi}{2})$ are real parameters. This distribution is characterized by $\langle e^{i \alpha} \rangle = S e^{i \tilde{\alpha}} = S \tilde{\lambda}$ and $\langle e^{i m \alpha} \rangle = 0$ for all $|m|>1$, i.e. only its first harmonic is non-zero. This means that $\langle \eta \rangle = S b \tilde{\lambda}$, while $\langle \eta^m \rangle = 0$ for all $|m|>1$. Putting this in the series expression above, we obtain a closed simple formula for $\bbZZ$:
\begin{equation} \bbZZ = (1 - b^2) \big(  e^{i\PP}  + S b \tilde{\lambda} e^{i2\PP} \big)  - S b \tilde{\lambda}^*\;.  \label{eq-walphasolution}  \end{equation}

As our final example, we examine non-identical oscillators with Lorentzian frequency distribution Eq.~\er{eq-lorentzian}. We assume the parameters $b$ and $\lambda$ to be constant (but since the oscillators are not synchronized prior to the kick, this situation is not trivial). We have:
\[ \begin{array}{lll}
  \bbZZ & =  \frac{1}{\eta} \sum_{n=0}^{\infty}  \eta^{n+1}  \big\langle {\s_0}^{n+1} \big\rangle - \eta^* \sum_{n=0}^{\infty}  \eta^n  \big\langle {\s_0}^n \big\rangle  &  \\[0.2cm]
        &   =  (\frac{1}{\eta} - \eta^*) \sum_{n=0}^{\infty}  \eta^n  {(Z_0)}_n  - \frac{1}{\eta} &
\end{array} \] 
where the quantities ${(Z_0)}_n = \langle {\s_0}^n \rangle$ are defined as in~\cite{OA}. As shown there, the collective parameters describing the emergent dynamics of this system lie on OA manifold (we discuss this in detail in Section~\ref{Sec-Time-evolution}), defined by:
\[  {(Z_0)}_n  = (Z_0)^n \;\;\; \mbox{for all} \;\;\; n>1 \;. \]
We assume that the departure from OA manifold due to the kicking is not significant, i.e. the identity above holds. This assumption gives:
\begin{equation} \bbZZ = (\frac{1}{\eta} - \eta^*) \sum_{n=0}^{\infty}  \eta^n  (Z_0)^n  - \frac{1}{\eta}   = \frac{Z_0 - \eta^*}{1 - \eta Z_0}  \label{eq-OAmanifold}  \end{equation}
which is the OA approximation for $\bbZZ$. Note that this is a direct generalization of the phase jump formula for a single oscillator Eq.~\er{eq-kickjump-s}.

%------------------------------------------------------------------------------------------------------------------------------------------------------------------------------------------

\section{Time evolution of the phase reset}  \label{Sec-Time-evolution}

In this Section we employ the Ott-Antonsen theory and estimate the additional phase shift due to the evolution after the kick. We construct a formula for fPRC $\Delta_\infty$ for a general Sakaguchi-Kuramoto ensemble.

\subsection{An illustrative example}
\begin{figure}[!ht]  \begin{center}
       \includegraphics[height=2.43in,width=3.41in]{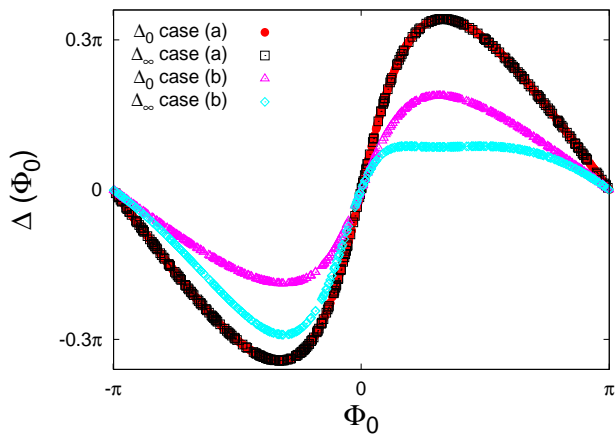}
  \caption{(Color online) PRC for an ensemble of two identical oscillators having $\om=1$, with two different kicking schemes. (a): $A_1 = A_2 = 0.18$, (b) $A_1 = 0.2, \; A_2 = 0$. 
      Both cases: $\beta=\frac{\pi}{3}$ and $\e=0.1$. Legend indicates various cases/curves.}  \label{fig-2} 
\end{center} \end{figure}

We begin by illustrating the concepts of pPRC and fPRC further. Consider the simplest nontrivial ensemble consisting of two oscillators, with dynamics given by Eq.~\er{eq-KSmodel} or \er{eq-KSmodel-s}. We apply two different versions of the kicking in the system's synchronized state: \\[0.1cm]
(a) both oscillators kicked with the equal strengths $A_1 = A_2 = 0.18$ \\[0.1cm]
(b) one oscillator kicked with $A_1=0.2$, the other left unperturbed ($A_2=0$) \\[0.1cm]
For both cases we numerically compute $\Delta_0$ and $\Delta_\infty$. Results are reported in Fig.~\ref{fig-2}: while in the case (a) we have a perfect agreement between $\Delta_0$ and $\Delta_\infty$, in the case (b) fPRC $\Delta_\infty$ significantly differs from the pPRC $\Delta_0$. 
The difference between the curves $\Delta_0$ and $\Delta_\infty$ in case (b) is not uniform, but it depends on the collective phase value. 

In the case (a) both oscillators receive equal kicks and hence the same immediate phase shift. This leaves the system in the synchronized state $\RR=\bbRR=1$, generating no transient relaxation: fPRC is given by the pPRC. In contrast to this, in the case (b) non-uniform kicking desynchronizes the system $\bbRR<\RR$, which generates an additional phase shift during the ensemble's relaxation (cf. Fig.~\ref{fig-1}). In the reminder of this Section we provide a theoretical framework for approximating the fPRC based on the knowledge of $\bbZZ$.

\subsection{Ott-Antonsen theory for non-identical oscillators} 
The Sakaguchi-Kuramoto ensemble of globally coupled identical oscillators can be exactly described by the Watanabe-Strogatz (WS) theory~\cite{watanabe}. However, the evolution of mean field in this theory is described through the constants of motion, that are determined from the initial conditions by rather complicated expressions. Recently, Ott and Antonsen found a simple ansatz that leads to a closed set of equations for the mean field~\cite{OA}. As shown in~\cite{PR}, this ansatz corresponds to a special choice of constants of motion in the WS theory. Such a dynamical state (defining the OA manifold) is eventually reached in an ensemble of non-identical oscillators with Lorentzian frequency distribution. In this paper we apply the OA theory by assuming that the deviations from OA manifold due to the kick are small (cf. derivation of the Eq.~\er{eq-OAmanifold}). We obtain simple (although approximate) analytical expressions, whose precision is shown through numerical results in Section~\ref{Sec-Numerics}.

We thus consider the general case of an ensemble with collective dynamics given by Eq.~\er{eq-KSmodel} or \er{eq-KSmodel-s}, and frequencies $\om$ picked from the Lorentzian distribution Eq.~\er{eq-lorentzian}. In the OA approximation, the evolution of the mean field $Z$ (in thermodynamic limit) is given by~\cite{OA}:
\begin{equation} \dot{Z} = i\om Z - \gamma Z + \frac{\varepsilon}{2} Z (e^{i\beta} - e^{-i\beta} |Z|^2)\;.  \label{eq-lorZequation} \end{equation}
After substituting $Z=R e^{i\Phi}$ this equation is decomposed into two real ODEs: 
\[ \begin{array}{lll}
 \dot{R} &=& \dfrac{\e \cos \beta}{2} R (1-R^2) - \gamma R \;, \\
 \dot{\Phi} &=& \omega + \dfrac{\e \sin \beta}{2} (1+R^2) \;.
\end{array} \]
The only stable fixed point for the first equation is $R_f = \sqrt{1 - \frac{2 \gamma}{\e \cos\beta}} \le 1$. The system achieves this final radius value only if $\e \cos\beta > 2\gamma$, and otherwise remains fully desynchronized with $R(t) \rightarrow 0$ (note that full synchronization $R_f=1$ occurs only for $\gamma=0$). For $R=R_f$ the collective phase rotates with frequency $\Omega = \omega + \e \sin\beta - \gamma \tan \beta$. Both equations above can be easily integrated, yielding:
\[  \begin{array}{lll}
R(t) &=  R_f  \left( 1 + \frac{R_f^2 - \RR^2}{\RR^2} e^{- ( \e \cos\beta - 2 \gamma ) \; t}  \right)^{-1/2} & \\   &  &  \\
\Phi(t) &= \PP + \omega t + \e \sin\beta \; t - \gamma \tan \beta \; t  + & \\[0.1cm]
\frac{\tan\beta}{2} & \ln \left( 1 + \frac{R_f^2 - \RR^2}{\RR^2} e^{- (\e \cos\beta - 2\gamma ) \; t} \right)  -  \tan\beta \; \ln \left( \frac{R_f}{\RR} \right) & 
\end{array} \]
where $\RR = R(0)$ and $\PP = \Phi (0)$. Consider again the original and kicked ensemble. The original system is at time $t=0$ in stationary state $\ZZ$, with radius $R_0$ (equals to final stationary radius $R_0 = R_f$) and phase evolution $\Phi (t) = \PP + \Omega t$. The kicked system starts at time $t=0$ from the initial state $\bbZZ = \bbRR e^{i\bbPP}$, with the time evolution given by the equation above. We consider the phase difference between two systems defined as $\Delta (t) = \big(\bbP - \Phi \big) (t)$ which reads:
\begin{equation} \begin{array}{l}
\Delta (t) = (\bbPP - \PP) +  \\[0.1cm]
\frac{\tan\beta}{2} \ln \left( 1 + \frac{\bar{R}_f^2 - \bbRR^2}{\bbRR^2} e^{- ( \e \cos\beta - 2 \gamma ) \; t} \right) + \tan\beta \; \ln \left( \frac{\bbRR}{\bar{R}_f} \right)\;. \end{array}  \label{eq-lordeltaTsolution}
\end{equation}
Coming back to the definitions of pPRC and fPRC from Section~\ref{Sec-Formulation}, we differ between two situations:
\begin{itemize}
 \item at $t=0$ last two terms in Eq.~\er{eq-lordeltaTsolution} cancel out, leaving   
                \[ \Delta (t=0) = \bbPP - \PP = \Delta_0 \]  
       which is the pPRC $\Delta_0$, consistently with what exposed in Section~\ref{Sec-Formulation}.
 \item as $t \rightarrow \infty$ the middle term on RHS of Eq.~\er{eq-lordeltaTsolution} vanishes at the limit, leaving 
    \begin{equation}  \begin{array}{ll}
          \Delta_\infty & =   \Delta_0 + \Delta_R   \;\;\;\;\;\;   \mbox{where we have defined}  \\[0.2cm]
          \Delta_R  & =   \tan\beta \; \ln \left( \dfrac{\bbRR}{\RR} \right)   =  \tan\beta \; \ln \left| \dfrac{\bbZZ}{Z_0}  \right| 
    \end{array}  \label{eq-deltaR} \end{equation}  
    (note that $\RR = R_f = \bar{R}_f$, as the final stationary $R$-value depends only on the ensemble parameters). 
\end{itemize}
The term $\Delta_R$ is what accounts for the additional phase shift due to system's relaxation back to stationary state. It depends on the nature of perturbation through $\frac{\bbRR}{\RR}$ and the dynamics of kicked system through $\beta$. For perturbations that leave $\RR=\bbRR$ (e.g. equal perturbation applied to all oscillators as in Fig.~\ref{fig-2} case (a)) and systems with $\beta=0$, the fPRC $\Delta_\infty$ is identical to pPRC $\Delta_0$. On the other hand, if $\beta \neq 0$ the rotation speed of the collective phase depends on the value of $R$. Hence, if the system's stationary radius is perturbed ($\RR \neq \bbRR$), an additional phase reset $\Delta_R$ is generated at the limit (cf. Fig.~\ref{fig-1} and Fig.~\ref{fig-2} case (b)), which depends on the kicking phase $\PP$.

If the details of the kick are known $\Delta_R$ can be calculated from Eq.~\er{eq-deltaR} and the analytical approximation for fPRC can be constructed. Alternatively, fPRC can be numerically approximated by summing the computed curves $\Delta_0$ and $\Delta_R$, even without knowing the parameters that specify the system and the kick. We stress again that Eq.~\er{eq-deltaR} is approximate, as it is based on the OA ansatz; its precision is related to the deviations from the OA manifold.

\subsection{Ott-Antonsen theory for identical oscillators} 
The equivalent results for the ensemble of identical oscillators can be easily recovered from what just exposed by taking $\gamma \rightarrow 0^+$. Since identical oscillators are always synchronized prior to the kicking, we also have 
$\RR = R_f = \bar{R}_f = 1$, and $\bbRR \le 1$. The phase difference equation Eq.~\er{eq-lordeltaTsolution} reduces to:
\[ \begin{array}{ll}
\Delta (t) = & (\bbPP - \PP) +  \\[0.1cm]
             & \frac{\tan\beta}{2} \ln \left( 1 + \frac{1 - \bbRR^2}{\bbRR^2} e^{- \e \cos\beta \; t} \right) + \tan\beta \; \ln \bbRR \; , \end{array}  \]
from which we obtain the formula for fPRC:
\[  \Delta_\infty = \Delta_0  + \tan\beta \; \ln \big| \bbZZ \big|\;. \]
The dynamics of the complex order parameter given by Eq.~\er{eq-lorZequation} only weakly depends on $\gamma$, and the limit $\gamma \rightarrow 0^+$ in this equation is not singular. However, the whole validity of the OA theory leading to Eq.~\er{eq-lorZequation} heavily depends on the spreading range of the frequencies $\gamma$, and the limit $\gamma \rightarrow 0^+$ appears to be singular in this respect (see discussion in \cite{OA,PR}).

In the next Section we show the numerical results confirming the theory just exposed.

%------------------------------------------------------------------------------------------------------------------------------------------------------------------------------------------

\section{Numerical results}   \label{Sec-Numerics}

In this Section we expose our numerical results, confirming our theoretical findings from previous Sections. For a given ensemble consisting of $N$ oscillators, we start from a random initial phase for each oscillator $\p \in [0,2\pi)$ and run the dynamics until the system reaches the stationary state. Then, a collective phase value $\PP$ is chosen at random (in the stationary regime) and the kick is applied (in accordance with Eq.~\er{eq-kickjump} for each oscillator). The phase reset is immediately measured, and the respective values of $\Delta_0 (\PP)$ and $\Delta_R (\PP)$ are recorded. The kicked system is then relaxed into a stationary state, when the final phase reset $\Delta_\infty (\PP)$ is measured with respect to the original system (at the same final time value). Repeating this procedure for many values of $\PP$ yields the numerical curves  $\Delta_0$, $\Delta_R$ and $\Delta_\infty$. These are compared with the corresponding analytical curves obtained as previously discussed. In the last paragraph we examine the phase-resetting in an ensemble of complex Stuart-Landau oscillators, showing the validity our results in a broader context.

\subsection{Small ensembles of identical oscillators}
\begin{figure}[!ht]  \begin{center}
       \includegraphics[height=2.43in,width=3.41in]{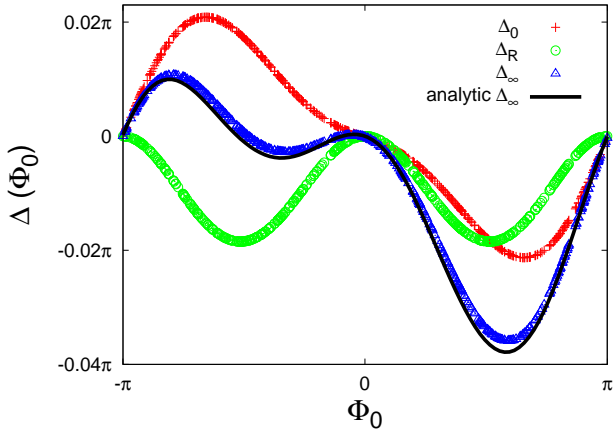}
  \caption{(Color online) Ensemble of $N=10$ identical oscillators, 3 kicked with a strength $A_1=0.05$, 4 kicked with a strength $A_2=-0.06$, and the remaining 3 unperturbed. 
           Parameters: $\alpha_k=0, \; \om=1, \; \e=0.1, \; \beta=\frac{2\pi}{7}$. Numerical and analytical curves are indicated in Legend.}  \label{fig-3} 
\end{center} \end{figure}

We start with an ensemble of $N=10$ identical oscillators with the kick implemented as follows: 3 oscillators are kicked with a strength of $A_1$, 4 are kicked with a strength of $A_2$, while the remaining 3 are left unperturbed. For this case $\bbZZ$ can be easily calculated using Eqs.\er{eq-defIPRC}\,\&\,\er{eq-kickjump-s}, which gives $\Delta_0$ and $\Delta_R$, and hence the fPRC $\Delta_\infty$. In Fig.~\ref{fig-3} we show the computed numerical curves $\Delta_0$, $\Delta_R$ and $\Delta_\infty$ (marked by different symbols), in comparison with the analytical fPRC $\Delta_\infty$ (thick curve).
The OA equation Eq.~\er{eq-lorZequation} is an approximation valid at the thermodynamic limit. Nevertheless, in Fig.~\ref{fig-3} analytical fPRC approximates the numerical fPRC with a good precision, despite small ensemble size. We considered many different deterministic kick realizations (through parameters $A$ and $\alpha$) for $N=10$ identical oscillators, and in general obtained a very good agreement between the numerical and analytical fPRC. Note that the magnitude of $\Delta_R$ is comparable to that of $\Delta_0$ -- the post-kick evolution of the phase reset is significant.

We now pick the
\begin{figure}[!ht]  \begin{center}
       \includegraphics[height=2.43in,width=3.41in]{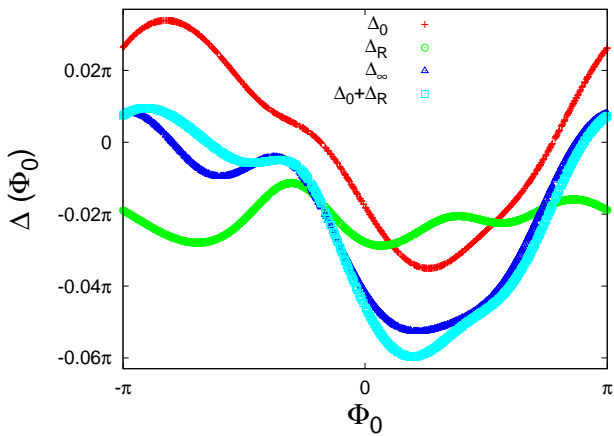}
  \caption{(Color online) Ensemble of $N=10$ identical oscillators with random kicks: strengths $A_k$ picked from a Gaussian distribution with mean 0 and standard deviation 0.1, $\alpha_k$-values picked uniformly from $[-\pi,\pi]$.
           Parameters: $\om=1, \; \e=0.1, \; \beta=\frac{2\pi}{7}$. Numerical curves shown as indicated in Legend.}  \label{fig-4} 
\end{center} \end{figure}
kicking parameters at random: $A_k$-values from a Gaussian distribution centered at $0$, and $\alpha_k$-values uniformly from interval $[-\pi,\pi]$. Numerically computed curves $\Delta_0$, $\Delta_R$ and $\Delta_\infty$ are reported in Fig.~\ref{fig-4} (same symbols are used). Since the details of the kick are now unknown, we are unable to construct the analytical approximations. However, it follows from the formula Eq.~\er{eq-deltaR} that one can sum numerically obtained $\Delta_0$ and $\Delta_R$ and approximate $\Delta_\infty$. We show this in Fig.~\ref{fig-4}: numerical curves $(\Delta_0 + \Delta_R)$ and $\Delta_\infty$ agree rather well, thus confirming the practical validity of the formula Eq.~\er{eq-deltaR}.

\subsection{Large ensembles of identical oscillators}
We now examine large ensembles approximating the thermodynamic limit through three examples discussed in Section~\ref{Sec-AnalyticalZ0}. We start with $N=1000$ identical oscillators and the kick implemented by uniformly picking the kicking
\begin{figure}[!ht] \begin{center}
  \includegraphics[height=4.78in,width=3.41in]{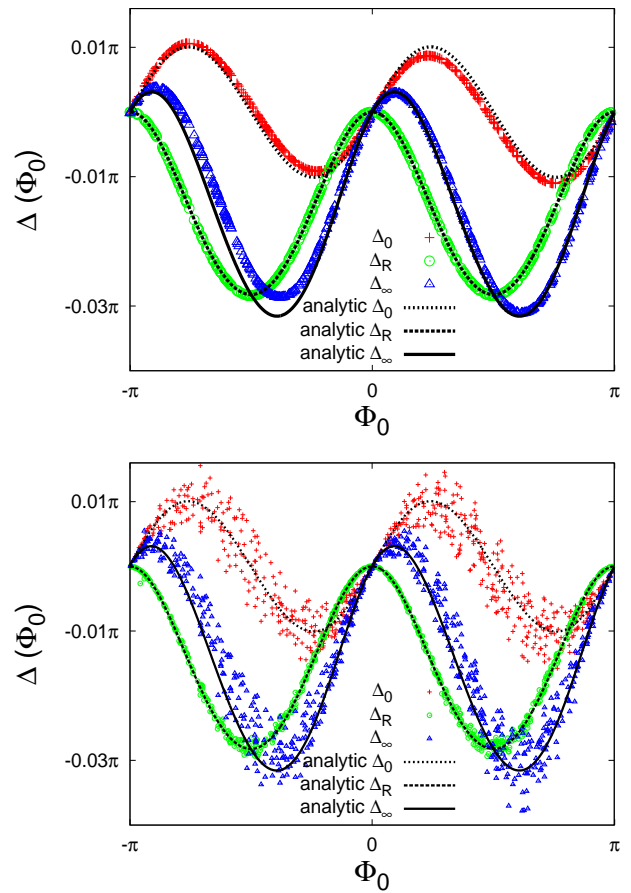}
  \caption{(Color online) Ensemble of $N=1000$ identical oscillators with kicking strengths $A$ picked from $A \in [-A_0,A_0]$ for $A_0=0.1$, with all $\alpha=0$, and $\om=1, \; \e=0.1, \; \beta=\frac{2\pi}{7}$. 
           All the curves are shown as indicated in the Legend. Top: a single realization of $A$-values. Bottom: 10 values taken from each of 60 runs with different realizations of $A$-values.}  \label{fig-5}
\end{center} \end{figure}
strengths $A$ from the interval $[-A_0,A_0]$, while setting $\alpha=0$ for all oscillators. We compute the numerical curves $\Delta_0$, $\Delta_R$ and $\Delta_\infty$, and construct the corresponding 'analytical' curves from Eq.~\er{eq-bbZZ-int} (quotes indicate that the integral Eq.~\er{eq-bbZZ-int} was solved numerically). Both numerical and analytical results are shown in Fig.~\ref{fig-5} (same symbols as previously). The top figure reports results of a run with a single realization of $A$-values, while in the bottom figure we show sampled data obtained from many different realization of $A$-values. For completeness, we now report all three analytical curves (differently dashed thick lines). There is a very good agreement between all numerical and analytical curves, even for a single realization of $A$-values, due to the large ensemble size. For the sampled curves, note that the theoretical fPRC almost perfectly approximates the averaged value of numerical fPRC. This further confirms the quality of the OA approximation. 

We now set all
\begin{figure}[!ht] \begin{center}
  \includegraphics[height=4.78in,width=3.41in]{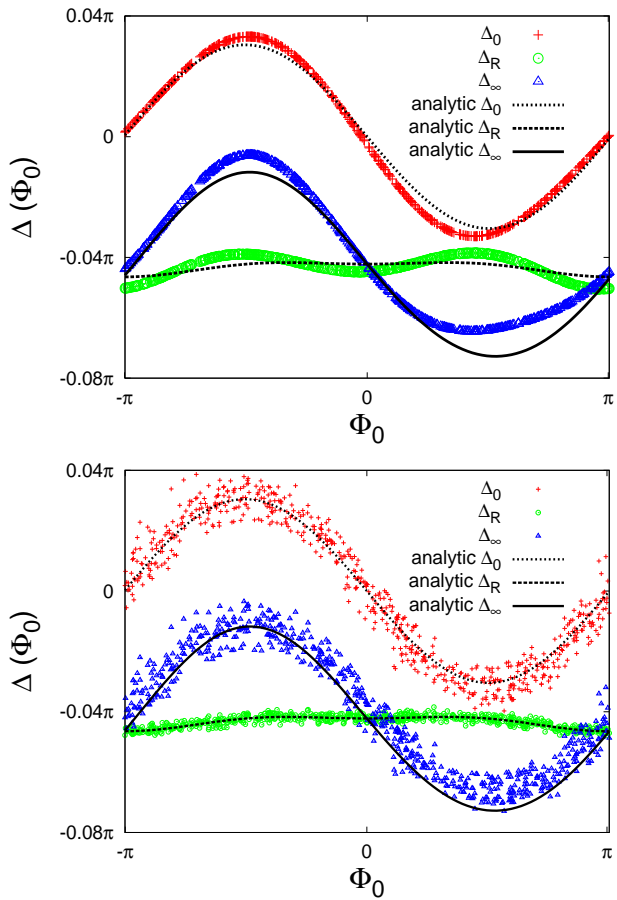}
  \caption{(Color online) Ensemble of $N=1000$ identical oscillators. All kicking strengths are fixed to $A_0=0.1$, while $\alpha$-values are picked from $[-\pi,\pi]$ via distribution 
            Eq.~\er{eq-walpha} for $S=0.15$ and $\tilde{\alpha}=\pi$. $\om=1, \; \e=0.1, \; \beta=\frac{2\pi}{7}$. All the curves are shown as indicated in the Legend. 
            Top: a single realization of $\alpha$-values. Bottom: 10 PRC-values taken from each of 60 runs with different realizations of $\alpha$-values.}   \label{fig-6}
\end{center} \end{figure}
the kicking strengths to constant $A=A_0$, while picking the $\alpha$-values from the interval $[-\pi,\pi]$ via the distribution Eq.~\er{eq-walpha}. We use Eq.~\er{eq-walphasolution} to test our numerical findings. Note that in opposition to the previous example, now we are having an \textit{entirely analytical} solution for the whole PRC-problem. The numerical and analytical results are reported in Fig.~\ref{fig-6}, where we again show a case of a single $\alpha$-values realization (top), and the sampled data from many different realization of $\alpha$-values (bottom). We have a very good agreement between all obtained curves, particularly in the sampled case. The agreement is however not perfect, and it can be used to estimate the precision of OA approximation.

\subsection{Large ensembles of non-identical oscillators}
We now examine an ensemble consisting of $N=1000$ non-identical oscillators with a Lorentzian frequency distribution Eq.~\er{eq-lorentzian}. 
\begin{figure}[!ht] \begin{center} 
  \includegraphics[height=4.78in,width=3.41in]{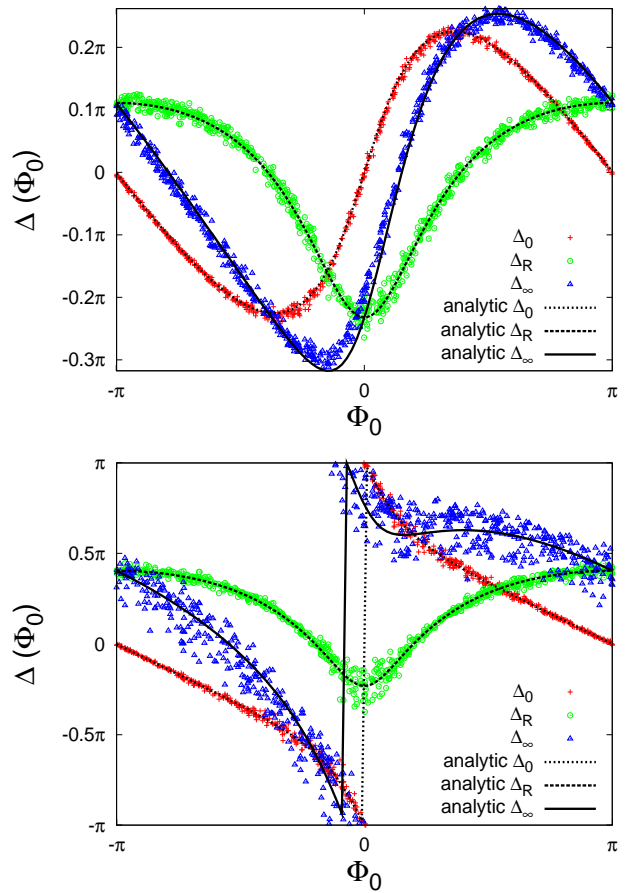}
  \caption{(Color online) Ensemble of $N=1000$ non-identical oscillators with Lorentzian frequency distribution Eq.~\er{eq-lorentzian} with $\tilde{\omega}=10$. All kicking strengths $A_0=0.1$ and $\alpha=0$.
           Other parameters: $\e=1, \; \beta=\frac{2\pi}{7}$. Top: $\gamma=0.2$, bottom: $\gamma=0.3$. All the curves are shown as indicated in the Legend. 
           Frequencies $\om_k$ are taken to be $\om_k = \gamma \tan (-\frac{\pi}{2} + \frac{k-0.5}{N}\pi)$.} \label{fig-7}
\end{center}  \end{figure}
All kicking strengths are set to $A=A_0$, with $\alpha=0$. Since the ensemble is not fully synchronized prior to the kicking, phase jumps of oscillators are in general different, depending on their phases value at the time of kick. We now use Eq.~\er{eq-OAmanifold} to obtain the analytical curves, assuming the kick does not remove the system significantly from the OA manifold. The results are shown in Fig.~\ref{fig-7} for two different values of Lorentzian width $\gamma$. We show only the cases of single kicking realizations (since they are always noisy). Again, a very good agreement between all the numerical and analytical curves is obtained, even in the case of $\gamma=0.3$ (bottom) where we observe a discontinuity in the pPRC and fPRC. This confirms the assumption of small departure from OA manifold due to kicking, used to obtain Eq.~\er{eq-OAmanifold}.

\subsection{Large ensembles of Stuart-Landau oscillators}
\begin{figure}[!ht] \begin{center} 
  \includegraphics[height=2.43in,width=3.41in]{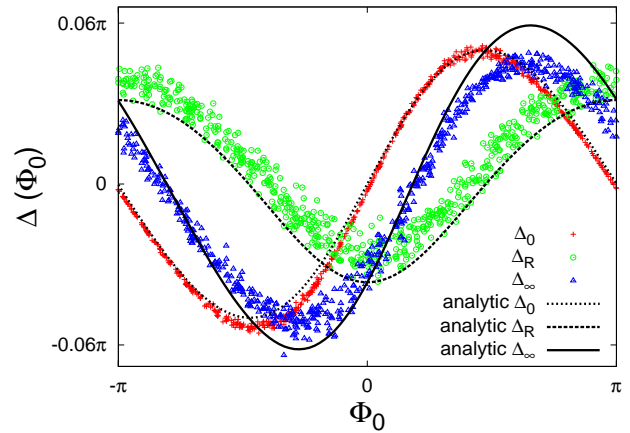}
  \caption{(Color online) Ensemble of $N=1000$ Stuart-Landau oscillators with Lorentzian frequency distribution $\tilde{\omega}=10, \; \gamma=0.2$ (cf. Fig.~\ref{fig-7}). All kicking strengths $A'_0=0.1,\; \alpha'=0$. 
           Other parameters: $\e=1,\; \beta=\frac{2\pi}{7},\; \xi=10$. All the curves are shown as indicated in the Legend (analytical ones are obtained using Eq.~\er{eq-OAmanifold} for $A=0.022$ 
           -- see text for details).} \label{fig-8}
\end{center}  \end{figure}

In the final part of this Section, we wish to illustrate the extent of the theory developed in this work. We consider an ensemble of $N=1000$ non-identical Stuart-Landau oscillators~\cite{kuramoto-book}, described by the complex amplitudes $w = r e^{i\p}$. The system's equation which reads:
\[ {\dot w} = \big(\xi + i \om - \xi |w |^2 \big) w  +  \e e^{i\beta} Z  - A' e^{-i \alpha'} \delta (t)  \]
can be decomposed into:
\[ \begin{array}{l}
{\dot r} = \xi (1 - r^2) r  +  \e R \cos (\Phi - \p + \beta)  \\[0.1cm]
\;\;\;\;\;\;\;\;\;\;\;\;\;\;\;\;\;\;\;\;\;\;\;\;\;\;\;\;\;\;\;\;\;\;\;\;\;\;\;\;\;\;\;\;\;\;\;\;\;\;  - A' \cos (\p + \alpha') \; \delta (t) \\[0.2cm]
{\dot \p} = \om  +  \e \frac{R}{r} \sin (\Phi - \p + \beta)  +  \frac{A'}{r} \sin (\p + \alpha') \; \delta (t)
\end{array} \]
where $Z = R e^{i\Phi} = \langle w \rangle$ is the mean field, through which the ensemble is coupled. For $\xi \gg 1$ the system's behavior resembles phase-oscillators since $r \approx 1$ after transients for all oscillators. This makes the non-kicked part of the phase equation above reduce to the Eq.~\er{eq-KSmodel}. The kick is implemented in analogy with the Section~\ref{Sec-Formulation}, but now acts according to:
\[ \begin{array}{l}
{\bar r}_0 =  \;\; | r_0 e^{i\p_0}  -  A' e^{i\alpha'}  |   \\[0.15cm]
{\bar\p}_0 = \arg (r_0 e^{i\p_0}  -  A' e^{i\alpha'})
\end{array} \]
In equivalence with the previous paragraph, we consider an ensemble with a single realization of the Lorentzian frequency distribution, and a uniform kick $A'=A'_0, \; \alpha'=0$. In Fig.~\ref{fig-8} we show the 
numerical curves, along with the analytical ones which are obtained from the Eq.~\er{eq-OAmanifold}, but using $A=0.022$ instead of $A=0.1$. Namely, the kick of a strength $A'_0=0.1$ described by the equations above produces different shift in the collective phase, roughly five times smaller than a kick of strength $A=0.1$ given by Eq.~\er{eq-kickjump} (since it affects both $r$ and $\p$). Nevertheless, there is a good agreement between numerical and analytical curves, despite having considered a completely different (complex, i.e. two-dimensional) class of oscillators. This suggests that the range of application of both OA theory and our phase-resetting theory extends beyond simple phase-oscillators.

%------------------------------------------------------------------------------------------------------------------------------------------------------------------------------------------

\section{Conclusions}  \label{Sec-Conclusions}

We constructed a theory of the phase resetting of the collective phase  for ensembles of Sakaguchi-Kuramoto oscillators interacting globally via a non-odd coupling function. By employing the Ott-Antonsen theory we showed that the final phase resetting curve can be recovered from the collective system's state $\bbZZ$, measured immediately after the perturbation. We confirmed our theory on a series of numerical examples involving small and large ensembles of identical and non-identical oscillators, using various kicking patterns and strengths. In particular, we showed an example where the entire PRC-problem was solved analytically (Eq.~\er{eq-walphasolution}). Finally, by investigating PRC of an ensemble of Stuart-Landau oscillators, we showed that our theory is valid beyond phase-oscillator models.

Our main result captured by formula Eq.~\er{eq-deltaR}, provides three methods of obtaining the final PRC $\Delta_\infty$:
\begin{itemize}
 \item \textit{numerically}, from (a long) run (dark/blue symbols in Figs.~\ref{fig-3}-\ref{fig-7}),
 \item \textit{analytically}, by calculating the curve $\Delta_\infty$ using Eqs.\er{eq-defIPRC}\,\&\,\er{eq-deltaR} (full thick curve in Figs.~\ref{fig-3}-\ref{fig-7}),
 \item \textit{semi-analytically}, by applying the Eq.~\er{eq-deltaR} to the numerically obtained curves $\Delta_0$ and $\Delta_R$ (squares in Fig.~\ref{fig-4}).
\end{itemize}
Our numerical results show a good agreement for all three methods on all examined examples.

Our findings suggest that the system's relaxation after the perturbation can significantly alter its final PRC. In agreement with~\cite{ko-ermen}, we found that the initial and the final phase resets coincide for odd interaction functions ($\beta=0$) or perturbations that leave the collective state unchanged ($\bbRR=\RR$). However, once the collective radius $R$ is perturbed, different rotation speeds of the collective phases of the original and kicked systems (proportional to $\sin \beta \neq 0$) induce an additional non-trivial phase reset at the limit.

It is worth noticing the case of $|\beta| = \frac{\pi}{2}$: here our theory no longer holds, since the mean field is zero $Z=0$, and therefore the collective phase is not well defined. This renders impossible to generally define the concept of phase-resetting. However, in this case the kick transiently induces $Z \neq 0$ contrasting the disordered behavior for $Z=0$ -- this situation can perhaps be used to model the dynamical behavior underlying the appearance of Event Related Potentials, which are of great current interest in neuroscience and psychology~\cite{luck}.

Our theory is valid for arbitrary Sakaguchi-Kuramoto-type ensembles of identical oscillators (e.g., a nonlinear coupling~\cite{PR1} can be easily incorporated), and ensembles of non-identical oscillators with Lorentzian frequency distribution. Its limitations are in principle given by the range of validity of OA theory~\cite{OA,PR,PR1}; nevertheless, our numerical results indicate that our analytical findings apply even for conditions that are formally not included in the OA theory, such as small ensembles (far from thermodynamic limit).

An important extension of our results regards the generalization on complex networks. A general network connection topology will generate a more complicated post-kick time evolution of the collective phase, which is not covered by the OA theory. Understanding evolution of the phase reset on complex networks is of high relevance for problems related to the reconstruction of the network structure from the experimental data. By investigating the properties of the empirical PRCs, one might be able to recover the connectivity patterns of the underlying network~\cite{achuthan1,timme}.

Another relevant question revolves around the stochastic periodic systems, and different methods of defining their PRCs~\cite{jusi}. A complete theory of stochastic PRCs will be of great help in constructing the PRC from periodic (but noisy) experimental data. Finally, since a PRC can be generally defined for any rhythmic system, our results can be extended to other types/models of oscillators, such as different models of neural cells.

%------------------------------------------------------------------------------------------------------------------------------------------------------------------------------------------

\acknowledgments Z.~L. acknowledges the support from DFG via project FOR868. Many thanks to M.~Rosenblum, A.~D\'iaz-Guilera, H.~Daido, G.~Bordyugov and H.~Kori for useful suggestions.


\begin{thebibliography}{} 
\bibitem{winfree-book}  A.~T.~Winfree, "The Geometry of Biological Time", Springer, New York (2001).
\bibitem{tass-book}  P.~A.~Tass, "Phase resetting in medicine and biology: stochastic modelling and data analysis", Springer, Berlin (2007).
\bibitem{granada} A.~Granada \textit{at al.}, "Phase Response Curves: Elucidating the Dynamics of Coupled Oscillators", in \textit{Methods in Enzymology}, \textbf{454}, Comp. Methods, part A (2009).
\bibitem{kuramoto-book}  Y.~Kuramoto, "Chemical Oscillations, Waves, and Turbulence", Dover, Mineola, New York (2003).
\bibitem{ermentrout-96}  G.~B.~Ermentrout, "Type I Membranes, Phase Resetting Curves, and Synchrony", Neural Comput. \textbf{8}, 979 (1996).
\bibitem{galan-ermen}  R.~F.~Gal\'an, G.~B.~Ermentrout and N.~N.~Urban, "Efficient Estimation of Phase-Resetting Curves in Real Neurons and its Significance for Neural-Network Modeling", Phys. Rev. Lett. \textbf{94}, 158101 (2005).
\bibitem{tateno}  T.~Tateno and H.~P.~C.~Robinson, "Phase Resetting Curves and Oscillatory Stability in Interneurons of Rat Somatosensory Cortex", Biophys. Jour. \textbf{92}, 2, 683 (2007).   
\bibitem{ramon}  J.~L.~Perez Velazquez \textit{at al.}, "Phase response curves in the characterization of epileptiform activity", Phys. Rev. E \textbf{76}, 061912 (2007).
\bibitem{achuthan1}  S.~Achuthan and C.~C.~Canavier, "Phase-Resetting Curves Determine Synchronization, Phase Locking, and Clustering in Networks of Neural Oscillators", J. Neuroscience \textbf{29}, 16, 5218 (2009).
\bibitem{preyer}  A.~J.~Preyer and R.~J.~Butera, "Neuronal Oscillators in Aplysia californica that Demonstrate Weak Coupling In Vitro", Phys. Rev. Lett. \textbf{95}, 138103 (2005).
\bibitem{ota}  K.~Ota, M.~Nomura and T.~Aoyagi, "Weighted Spike-Triggered Average of a Fluctuating Stimulus Yielding the Phase Response Curve", Phys. Rev. Lett. \textbf{103}, 024101 (2009).
\bibitem{petri}  B.~Petri and M.~Stengl, "Phase Response Curves of a Molecular Model Oscillator: Implications for Mutual Coupling of Paired Oscillators", J. Biol. Rhythms \textbf{16}, 2, 125 (2001).
\bibitem{gonzo}  H.~Gonz\'alez, H.~Arce and M.~R.~Guevara, "Phase resetting, phase locking, and bistability in the periodically driven saline oscillator: Experiment and model", Phys. Rev. E \textbf{78}, 036217 (2008).
\bibitem{jeff} E.~Brown, J.~Moehlis and P.~Holmes, "On the Phase Reduction and Response Dynamics of Neural Oscillator Populations", Neural Comput. \textbf{16}, 673 (2004).
\bibitem{croisier} H.~Croisier, M.~R.~Guevara and P.~C.~Dauby, "Bifurcation analysis of a periodically forced relaxation oscillator: Differential model versus phase-resetting map", Phys. Rev. E \textbf{79}, 016209 (2009).
\bibitem{prk-book}  A.~Pikovsky, M.~Rosenblum and J.~Kurths, "Synchronization. A Universal Concept in Nonlinear Sciences" Cambridge University Press, Cambridge (2004).
\bibitem{acebron} J.~A.~Acebr\'on, "The Kuramoto model: A simple paradigm for synchronization phenomena", Rev. Mod. Phys. \textbf{77}, 137–185 (2005).
\bibitem{arenas}  A.~Arenas \textit{at al.}, "Synchronization in complex networks", Phys. Reports \textbf{469}, 3, 93 (2008).
\bibitem{kawamura}  Y.~Kawamura \textit{at al.}, "Collective Phase Sensitivity", Phys. Rev. Lett. \textbf{101}, 024101 (2008).
\bibitem{kori}  H.~Kori \textit{at al.}, "Collective-phase description of coupled oscillators with general network structure", Phys. Rev. E \textbf{80}, 036207 (2009).
\bibitem{ko-ermen}  T.~W.~Ko and G.~B.~Ermentrout, "Phase-response curves of coupled oscillators", Phys. Rev. E \textbf{79}, 016211 (2009).
\bibitem{talathi}  S.~S.~Talathi \textit{at al.}, "Predicting synchrony in heterogeneous pulse coupled oscillators", Phys. Rev. E \textbf{80}, 021908 (2009).
\bibitem{achuthan2}  S.~Achuthan and C.~C.~Canavier, "Phase response curves determine network activity of all to all networks of pulse coupled oscillators", BMC Neurosc. \textbf{9}, 1, 135 (2008).
\bibitem{kralemann}  B.~Kralemann \textit{at al.}, "Phase dynamics of coupled oscillators reconstructed from data", Phys. Rev. E \textbf{77}, 066205 (2008).
\bibitem{timme}  M.~Timme, "Revealing Network Connectivity from Response Dynamics", Phys. Rev. Lett. \textbf{98}, 224101 (2007).
\bibitem{sakaguchi}  H.~Sakaguchi and Y.~Kuramoto, "A Soluble Active Rotater Model Showing Phase Transitions via Mutual Entertainment", Prog. Theor. Phys. \textbf{76}, 576 (1986).
\bibitem{watanabe}  S.~Watanabe and S.~H.~Strogatz, "Constants of motion for superconducting Josephson arrays", Physica D \textbf{74}, 197 (1994). 
\bibitem{OA}  E.~Ott and T.~M.~Antonsen, "Low dimensional behavior of large systems of globally coupled oscillators", CHAOS \textbf{18}, 037113 (2008); E.~Ott and T.~M.~Antonsen, "Long time evolution of phase oscillator systems",
              CHAOS \textbf{19}, 023117 (2009).
\bibitem{PR}  A.~Pikovsky and M.~Rosenblum, "Partially Integrable Dynamics of Hierarchical Populations of Coupled Oscillators", Phys. Rev. Lett. \textbf{101}, 264103 (2008); A.~Pikovsky and M.~Rosenblum, 
             "Partially integrable dynamics of ensembles of nonidentical oscillators", Preprint: arXiv:1001.1299 (2010).
\bibitem{luck}  S.~J.~Luck, "An Introduction to the Event-Related Potential Technique", MIT Press, Cambridge (2005).
\bibitem{PR1}  M.~Rosenblum and A.~Pikovsky, "Self-Organized Quasiperiodicity in Oscillator Ensembles with Global Nonlinear Coupling", Phys. Rev. Lett. \textbf{98} 064101 (2007); A.~Pikovsky and M.~Rosenblum, 
              "Self-organized partially synchronous dynamics in populations of nonlinearly coupled oscillators", Physica D \textbf{238}, 27 (2009).
\bibitem{jusi}  J.~T.~C.~Schwabedal and A.~Pikovsky, "Effective phase dynamics of noise-induced oscillations in excitable systems", Phys. Rev. E \textbf{81}, 046218 (2010).
\end{thebibliography}
\end{document}